\documentclass[final,times]{elsarticle}
\usepackage{graphicx}
\usepackage[T1]{fontenc} 
\usepackage[normalem]{ulem}
\usepackage[dvipsnames]{xcolor}
\usepackage{float}
\usepackage{soul}

\journal{Physics Letters  B}

\begin{document}
\begin{frontmatter}
\title{Cosmic acceleration in entropic cosmology.} 
\author{J. Chagoya$^{a}$}
\ead{javier.chagoya@fisica.uaz.edu.mx}
\author{I. D\'iaz-Salda\~na$^{a}$}
\ead{isaacdiaz@fisica.uaz.edu.mx}
\author{J. C. L\'opez-Dom\'inguez$^{a,b}$}
\ead{jlopez@fisica.uaz.edu.mx}
\author{M. Sabido$^{b}$}
\ead{msabido@fisica.ugto.mx}
\address{$^a$Unidad Acad\'emica de F\'isica, Universidad Aut\'onoma de Zacatecas,\\ Calzada Solidaridad esquina con Paseo a la Bufa S/N C.P. 98060, Zacatecas, M\'exico.}
\address{$^b$Departamento de F\'{\i}sica de la Universidad de Guanajuato,\\
 A.P. E-143, C.P. 37150, Le\'on, Guanajuato, M\'exico.}

\begin{abstract}
In this paper we study  the viability of an entropic cosmological model. The effects of entropic gravity are derived from a modified entropy-area relationship  with a volumetric entropy term. This model describes a late time limit cosmic acceleration, whose origin is related to a volumetric term in the entropy. Moreover, we analyze the phenomenological implications of the entropic model using the Supernovae {\it Pantheon} compilation  and the observational Hubble parameter data to  find consistency with cosmological observations.
Finally, we show the equivalence between the entropic model and  a brane world cosmological model, by means of an effective geometrical construction.

\end{abstract}
\end{frontmatter}

\section{Introduction} \label{Int}
The verification of the current acceleration of the Universe opened a new avenue of research in cosmology. The best model is the standard model of cosmology, known as  lambda cold dark matter ($\Lambda CDM$). This model  is based on the existence of matter that possesses anomalous physical properties. In particular, conjectures the existence of  a cosmological constant $\Lambda$ as the 
source of the late time acceleration. Although there is good phenomenological agreement with observations,  there are serious theoretical concerns that need to be understood \cite{weinberg,lambda}. 

A different approach is to assume that this  acceleration is a consequence of a modified theory of gravity, several ideas have been pursued \cite{padilla1,padilla2,Horndeski,18,Od1,Od2}.  One possibility, is to consider gravity as an emergent phenomena \cite{Jacobson:1995ab}. The renewed interest in this idea started with Verlinde \cite{Verlinde:2010hp}, where the author claims that Newtonian gravity is an  emergent force. Moreover, modifications to  gravity can be induced by changing  the entropy-area relation. In this context, the possibility of a common origin of dark matter  and dark energy was proposed \cite{Verlinde2,odin1}. 
The dark matter predictions for this theory have been put to test in several works \cite{niz,test2, isaac1}. 
Following these lines of reasoning, a connection with dark energy was explored. More precisely, an entropic origin to the cosmological constant was proposed. This is achieved by considering a cosmological model that is derived from Clausius relation and a modified entropy-area relationship. The modified Friedmann equations in the limit $t\to \infty$ and $\rho\to 0$, gives a de Sitter Universe. This allowed to define an effective cosmological constant \cite{isaac2} and is related to the modification to Hawking-Bekenstein entropy. Considering the effective cosmological constant at $t\to \infty$, points that at present times this model describes an accelerating universe, and hence it should be equivalent to a cosmological model with a dark energy component. 

The purpose of this paper is {two fold. The first goal is to establish the viability of this model by adjusting the parameters using cosmological data and
comparing it with the $\Lambda CDM$ model. We find a time varying dark energy component that causes the cosmic acceleration and find the effective dark energy parameter $\omega_{eff}$ for the barotropic equation of state. Because introducing a volumetric term on the entropy can be considered as physically unmotivated, the second goal is to establish a physical understanding of the entropic model. We find an equivalence of the entropic cosmological model with the Dvali-Gabadadze-Porrati (DGP) cosmological model in \cite{deffayet}, which emerges in the brane worlds framework. Exploiting this equivalence we can relate the free parameter of our model with the characteristic distance scale of the DGP cosmology.}

This paper  is arranged as follows, in section \ref{eg} we review the derivation of the modified Friedmann equations  as well as some theoretical implications, mainly the de Sitter late time behavior and therefore the effective cosmological constant. In section \ref{cosmo}, we show that the model satisfies the cosmological tests and therefore a viable. 
Finally, section \ref{conclusiones} is devoted to  concluding remarks.

 \section{The modified Friedmann equation from the new entropy-area relationship}\label{eg}
Let us start by reviewing the derivation of the modified Friedmann equations based on the application of the Clausius relation on the apparent horizon of the FRW universe \cite{Cai:2005ra,cai}.

First, we consider a FRW universe described by the metric
\begin{equation}
ds^{2}=-dt^{2}+a^{2}(t)\left( \frac{dr^{2}}{1-\kappa r^{2}}+r^{2}d{\Omega}^{2}\right).
\end{equation}
The apparent horizon of the FRW universe is defined through the condition $h^{ab}\partial_{a}\tilde{r}\partial_{b}\tilde{r}=0$, where $\tilde{r}=a(t)r$ and $h^{ab}$ is identified by writing the metric as $ds^{2}=h_{ab}dx^{a}dx^{b} +\tilde{r}^{2}d\Omega^{2}
$. This condition  leads us {to} the radius of the apparent horizon
\begin{equation}
\tilde{r}^2_{A}=\left(H^2+\kappa/a^2\right)^{-1}.
\end{equation} 
To apply the Clausius relation we use the temperature and entropy on the apparent horizon \cite{Cai:2008gw}. For this work, we assume that the temperature is $T=1/2\pi \tilde{r}_A$ and the entropy is given by
\begin{equation}\label{entropy}
S(A)=\frac{A}{4G}+\epsilon\left(\frac{A}{4G}\right)^{3/2},
\end{equation}
where $A$ is the area of the apparent horizon  and $\epsilon$ is a free parameter of the model. We will use a modified entropy-area relationship that includes a volumetric contribution. The volumetric term has been obtained in loop quantum gravity \cite{livine} and also in the study of SUSY black holes \cite{isaac1}. Moreover, 
in a previous work \cite{isaac2}  it is inferred that the inclusion of a volumetric\footnote{ The
volumetric dependence is typically related to  degrees of
freedom in ordinary quantum field theories.} term is related to the late time acceleration of the Universe.

Using the energy-momentum tensor for a perfect fluid of energy density $\rho$ and
pressure $P$, the Clausius relation on the apparent horizon of the FRW universe gives
\begin{equation}\label{clausius}
4\pi \tilde{r}_{A}^{2}(\rho+P)H\tilde{r}_{A}dt=
\frac{1}{8\pi G}\frac{1}{\tilde{r}_{A}}\left[1+\frac{3\epsilon}{\sqrt{2G}}A^{1/2} \right] dA. 
\end{equation}{From} the time derivative of $A$ and the standard continuity equation {$\dot{\rho}+3H(p+\rho)=0$}, we arrive (after solving for $ H^2+\kappa /a^2$) to the modified Friedmann equation
\begin{equation}\label{mfe}
H^2+\frac{\kappa}{a^2}=\frac{8\pi G}{3}\rho+\frac{{9\pi}}{2G}\epsilon^2  
\pm \frac{1}{2}\sqrt{\frac{{81\pi^2}}{G^2}\epsilon^4+96\pi^2\epsilon^2 \rho }\quad.
\end{equation}

As our interest is late time cosmic acceleration,  we  consider the positive\footnote{The solution with the negative sign of the square root, does not present  cosmic acceleration in the limit $\rho\to 0$. Moreover, in this limit, the terms with $\epsilon$ cancel out, and there is no contribution from the volumetric term of the entropy. For these reasons, we discard this branch of the solution.} sign on the square root, as this branch of the solution gives  a de Sitter Universe for $t\to \infty$ and $\rho\to 0$.
Solving Eq.(\ref{mfe}) for a perfect fluid with a barotropic equation of state $P=\omega \rho$ and a flat FRW Universe, gives an implicit expression for the scale factor.
In the early-time limit, the behaviour of the scale factor is the same as the one derived from the usual Friedmann equation for a perfect fluid with a barotropic equation of state. However, the most important feature of this model is that in the late time limit, the scale factor exhibits an exponential growth modulated by the effective cosmological constant 
\begin{equation}
 \Lambda_{eff}=\frac{27\pi}{G}\epsilon^{2}.\label{lambda}
\end{equation}
Interestingly, this late-time exponential behaviour is provided by the volumetric correction term in Eq.(\ref{entropy}). Finally, it is important to emphasize that our model does not have a priori a cosmological constant, or a dark energy component of any kind.

\section{Phenomenological Tests }\label{cosmo}
In the previous sections we presented an entropic origin for the dark energy component that causes cosmic acceleration. More precisely, that the dark energy component is a consequence of the volumetric part on the entropy-area relationship.

In order to proceed with phenomenological tests of our model, it is suitable to express the modified Friedmann equations in terms of density parameters. First, let us note that Eq.(\ref{mfe}) can be expressed as 
\begin{equation}\label{despejado}
    H^2+\frac{\kappa}{a^2}=\frac{8\pi G}{3}\rho+\frac{1}{2}\frac{9\pi}{G}\epsilon^2+\sqrt{\frac{1}{4}\left(\frac{9\pi}{G}\epsilon^2\right)^{2}+\left(\frac{9\pi}{G}\epsilon^2\right)\left(\frac{8\pi G}{3}\rho\right)}\quad.
\end{equation}
This equation encourages us to define the parameter associated to $\epsilon$ as
\begin{equation}
\Omega_{\epsilon 0}\equiv\frac{1}{2H_{0}^2}\frac{{9\pi}}{G}\epsilon^{2}.
\end{equation}
Furthermore, the parameters associated to matter and spatial curvature are defined as usual,
\begin{equation}\label{epsilonparameter}
\Omega_{M 0}=\frac{8\pi G}{3H_{0}^{2}}\rho_{0},\quad \Omega_{\kappa 0}=-\frac{\kappa}{a_{0}^2H_{0}^2}~. 
\end{equation}
After considering non-relativistic matter for which $\rho=\rho_0 (1+z)^3$ where $z$ is the redshift  $1+z=1/a$, Eq.~(\ref{despejado}) can be written in terms of the above parameters as
\begin{equation}\label{friedred}
H^{2}(z)=H_{0}^{2}\left[ \Omega_{M0}(1+z)^{3}+\Omega_{\kappa 0}(1+z)^{2}+\Omega_{\epsilon 0}+ \sqrt{\Omega_{\epsilon 0}^{2}+2\Omega_{\epsilon 0}\Omega_{M 0}(1+z)^{3}}  \right].
\end{equation}
We can compare this equation with the usual Friedmann equation 
\begin{equation}\label{usual}
    H^{2}(z)=H_{0}^{2}\left\{\Omega_{\kappa 0}(1+z)^{2}+\Omega_{M 0}(1+z)^{3}+\Omega_{X 0}(1+z)^{3\left(1+\omega_{X}\right)}\right\},
\end{equation}
where $\Omega_{X 0}$ represents a dark energy component with equation of state parameter given by $\omega_{X}$. Usual cosmological constant is mimicked by $\omega_{X}=-1,$ in this case $\Omega_{X0}=\Omega_{\Lambda}$. For redshift $z=0$ the modified Friedmann equation  yields
\begin{equation}
\Omega_{M 0}+\Omega_{\kappa 0}+\Omega_{\epsilon 0}+ \sqrt{\Omega_{\epsilon 0}^{2}+2\Omega_{\epsilon 0}\Omega_{M 0}}=1,\label{modconstr}
\end{equation}
which is to be recognized as our modified Friedmann constraint for the density parameters. This constraint clearly differs from the usual $\sum_{i}\Omega_{i 0}=1$, obtained from Eq.(\ref{usual}). We can further rewrite  Eq.~(\ref{friedred}) in a simpler form by noting that
\begin{equation}
   \Omega_{M0}(1+z)^{3}+\Omega_{\epsilon 0}+ \sqrt{\Omega_{\epsilon 0}^{2}+2\Omega_{\epsilon 0}\Omega_{M 0}(1+z)^{3}} =\frac{1}{2}\left[\sqrt{\Omega_{\epsilon 0}}+\sqrt{\Omega_{\epsilon 0}+2\Omega_{M 0}(1+z)^{3}}\right]^{2},
\end{equation}
therefore we have
\begin{equation}\label{Hentropic}
H^{2}(z)=H_{0}^{2}\left\{ \Omega_{\kappa 0}(1+z)^{2}+ \frac{1}{2}\left[\sqrt{\Omega_{\epsilon 0}}+\sqrt{\Omega_{\epsilon 0}+2\Omega_{M 0}(1+z)^{3}}\right]^{2} \right\},
\end{equation}
and the modified Friedmann constraint Eq.(\ref{modconstr}) becomes
\begin{equation}\label{omegaentropic}
\Omega_{\kappa 0}+\frac{1}{2}\left(\sqrt{\Omega_{\epsilon 0}}+\sqrt{\Omega_{\epsilon 0}+2\Omega_{M 0}}\right)^{2}=1.
\end{equation}

With the results obtained so far, we can 
estimate a range of parameters where the model is compatible with background cosmology. 
{As a first approach towards phenomenological constraints, }
let us fit the parameters of our model
by comparing with two sets of observational data and minimizing the $\chi^2$ estimator of goodness of fit,
\begin{equation}\label{chi}
\chi_{O}^2=\sum_{i=1}^{N} \left(\frac{O_{th}(z_i, \Theta)-O_{obs}(z_i)}{\sigma_{obs}^{i}} \right)^2,
\end{equation}
where $O$ is the physical quantity under consideration, with theoretical predictions $O_{th}$ depending on certain parameters $\Theta$, and reported observational values $O_{obs}$ with corresponding uncertainties $\sigma_{obs}$, and $N$ is the number of data points. Although there are observational constraints on DGP reported in the literature~\cite{dgpconstraints}, in order to get a self-contained analysis with
more recent observations we constrain the model with 
the following data sets:
\begin{itemize}
\item {\it Supernovae 1a}. (SN Ia) Standard candles, 
such as type 1a supernovae, are particularly
useful for fixing the background cosmology
in theories of modified gravity. We use the 
{\it Pantheon} compilation~\cite{Pan-STARRS1:2017jku}, which consists of 1048
supernovae in the range $0.01<z<2.3$. The data reported in~\cite{Pan-STARRS1:2017jku}
consists of apparent magnitudes $m_b$, which are related to luminosity distance in megaparsecs by
\begin{equation}
m_b - M = 5 \log_{10}\left(d_{L}/{\rm Mpc}\right) + 25,
\end{equation}
where $M$ is the absolute B-band magnitude of a fiducial SN Ia. Using this relation, assuming a spatially flat universe, and using Eq.(\ref{Hentropic}), Eq.(\ref{omegaentropic}) and the luminosity distance 
\begin{equation}\label{distluminosa}
d_{L}= (1+z)H_0\int_{0}^{z}\frac{dz'}{H(z')},
\end{equation}
we construct the theoretical predictions for $m_b$, depending on the parameter $\Omega_{M 0}$, and we minimize Eq.(\ref{chi}) for $O = m_b$ and $N=1048$. The parameter $H_0$ is not included in the minimization of $\chi^2$, instead, it is determined as $H_0 = 71.4 $ in the region $z<0.05$ using the linear Hubble relation.
\item {\it Hubble parameter data.} (OHD) We use the
compilation provided by~\cite{Yu:2017iju}, which consists of 36 data points for $H(z)$, determined using cosmic chronometers and baryon acoustic oscillations, in the range $z<2.36$~\cite{Simon:2004tf,Stern:2009er,Moresco2012,Zhang:2012mp,BOSS:2014hwf,BOSS:2013igd,Moresco:2015cya,Moresco:2016mzx,BOSS:2016wmc,Ratsimbazafy:2017vga}. We compare this data to the predictions
of Eq.(\ref{Hentropic}) together with Eq.(\ref{omegaentropic}) and $\Omega_{\kappa 0} = 0$,
obtaining the parameters $(H_0, \Omega_{M 0})$ that minimize the $\chi^2$ estimator, Eq.(\ref{chi}), with $O = H$ and $N=36$.
\end{itemize}

\begin{figure}
\begin{center}
\includegraphics[width=.51\textwidth]{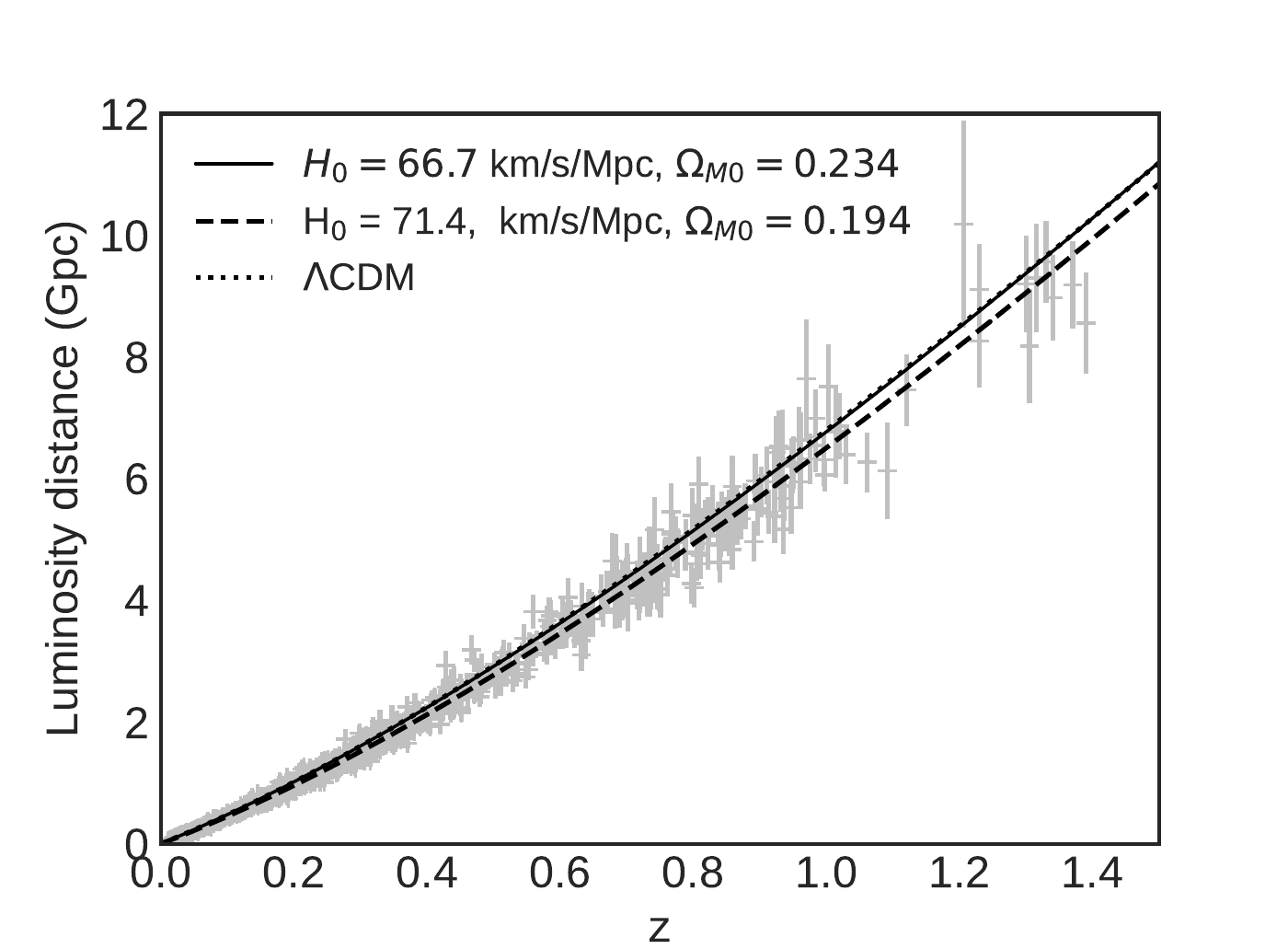}\includegraphics[width=.51\textwidth]{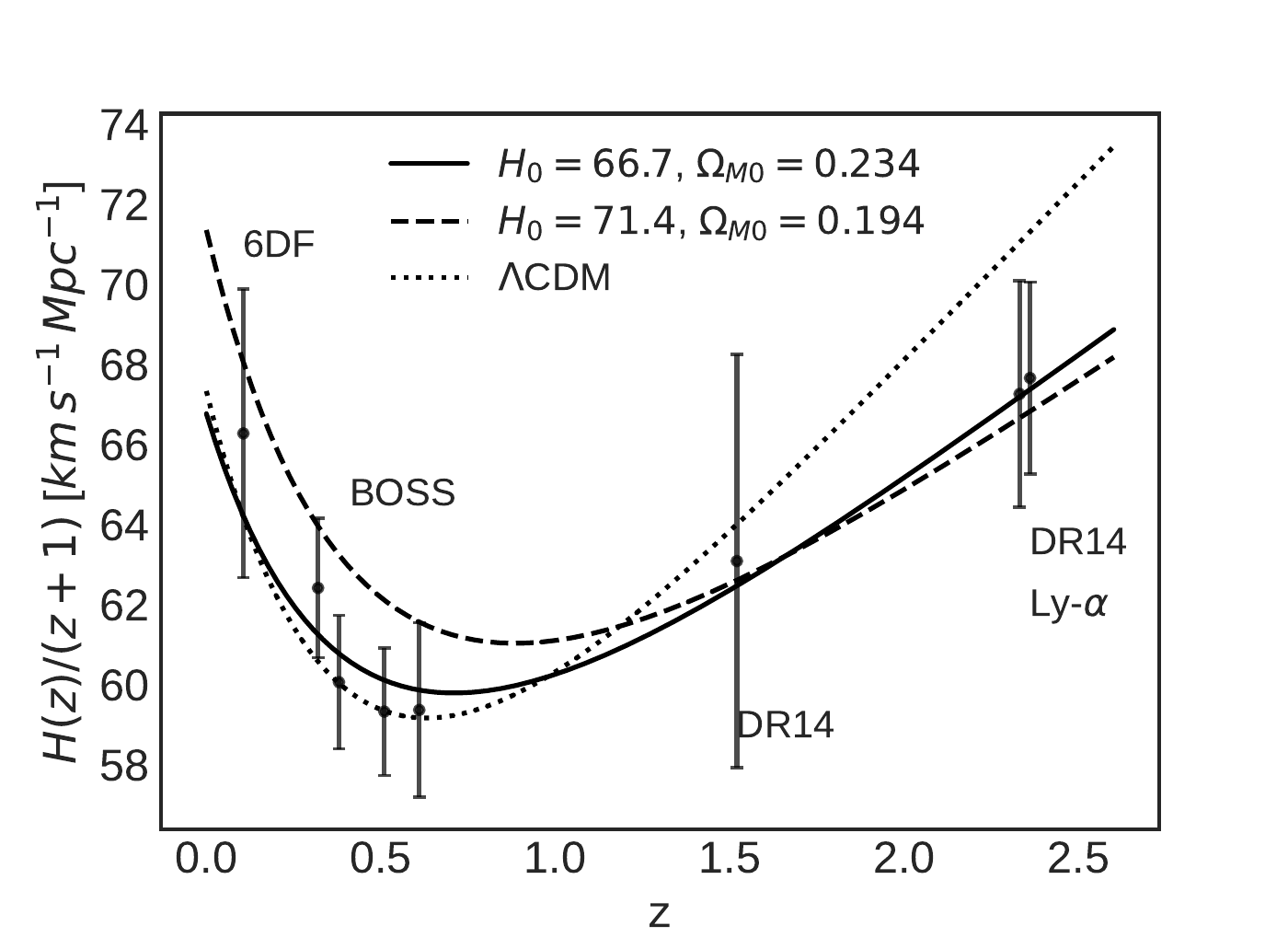}
\caption{Luminosity distance for SN Ia in the {\it Pantheon} compilation (left) and some sample points of $H(z)$ in the compilation~\cite{Yu:2017iju} (right). The curves in each plot correspond to the predictions of $\Lambda$CDM (dotted), the entropic model Eq.(\ref{Hentropic}) fitted to SN Ia (dashed) and the
entropic model fitted to evolution of $H(z)$ (solid).}
\label{figdata}
\end{center}
\end{figure}

The parameters for each of the data sets described above {are obtained  by minimization of $\chi^2$ using the L-BFGS-B algorithm, with tolerances of order $10^{-3}$ for SN Ia and $10^{-4}$ for OHD. For smaller tolerances the changes in the results for the best fits of the parameters are negligible. Furthermore, we estimate the uncertainties in these parameters by using the inverse Hessian, finding values of order $10^{-3}$ in each case.} The best fits obtained are presented in Table~\ref{tab:chisqr}, together with their reduced $\chi^2$. 
\begin{table}
  \centering  \begin{tabular}{@{}c|ccc}
        & $H_0$ & $\Omega_{M0}$  & $\chi_{red}^2$ \\ \hline
        SN Ia & 71.4 & $0.194$ &  $0.93$ \\  
        OHD & $66.7$ & $0.234$ & $0.48$  \\  
    \end{tabular}
    \caption{Best fit values and $\chi^2$ tests for the model described by Eq.(\ref{Hentropic}) with $\Omega_\kappa 0 = 0$, contrasted against supernovae 1a and Hubble parameter data. The small value of $\chi^2$ for OHD is a consequence of the large uncertainties in these measurements.}\label{tab:chisqr}
\end{table}
In Fig.~(\ref{figdata}) we show luminosity distance of SN Ia in the \textit{Pantheon} compilation (left panel) and the evolution of the Hubble parameter (right panel). In each plot, we include
as a reference the predictions of $\Lambda$CDM, and the predictions of our model for two sets of parameters: one corresponds to the best fit to $m_b$ described in the first item above (dashed lines), and the other is the best fit to $H(z)$ described in the second item (solid lines). Although the parameter used in the optimization is $\Omega_{M0}$, the value for the parameter $\Omega_{\epsilon 0}$ can be obtained from the constraint Eq.(\ref{omegaentropic}), i.e. we obtain two sets of values $(\Omega_{M 0}, \Omega_{\epsilon 0})$ which we will use in all the subsequent analysis {as references for exploring the parameter space of the theory. More rigorous bounds on the parameters of the model may be obtained by a joint analysis of multiple data sources and different statistical methods, such as Monte Carlo sampling. This will be explored in a future work. For now, we continue the analysis of the cosmological model in the space of parameters suggested by the optimization described above. }

\begin{figure}
\begin{center}
\includegraphics[scale=.69]{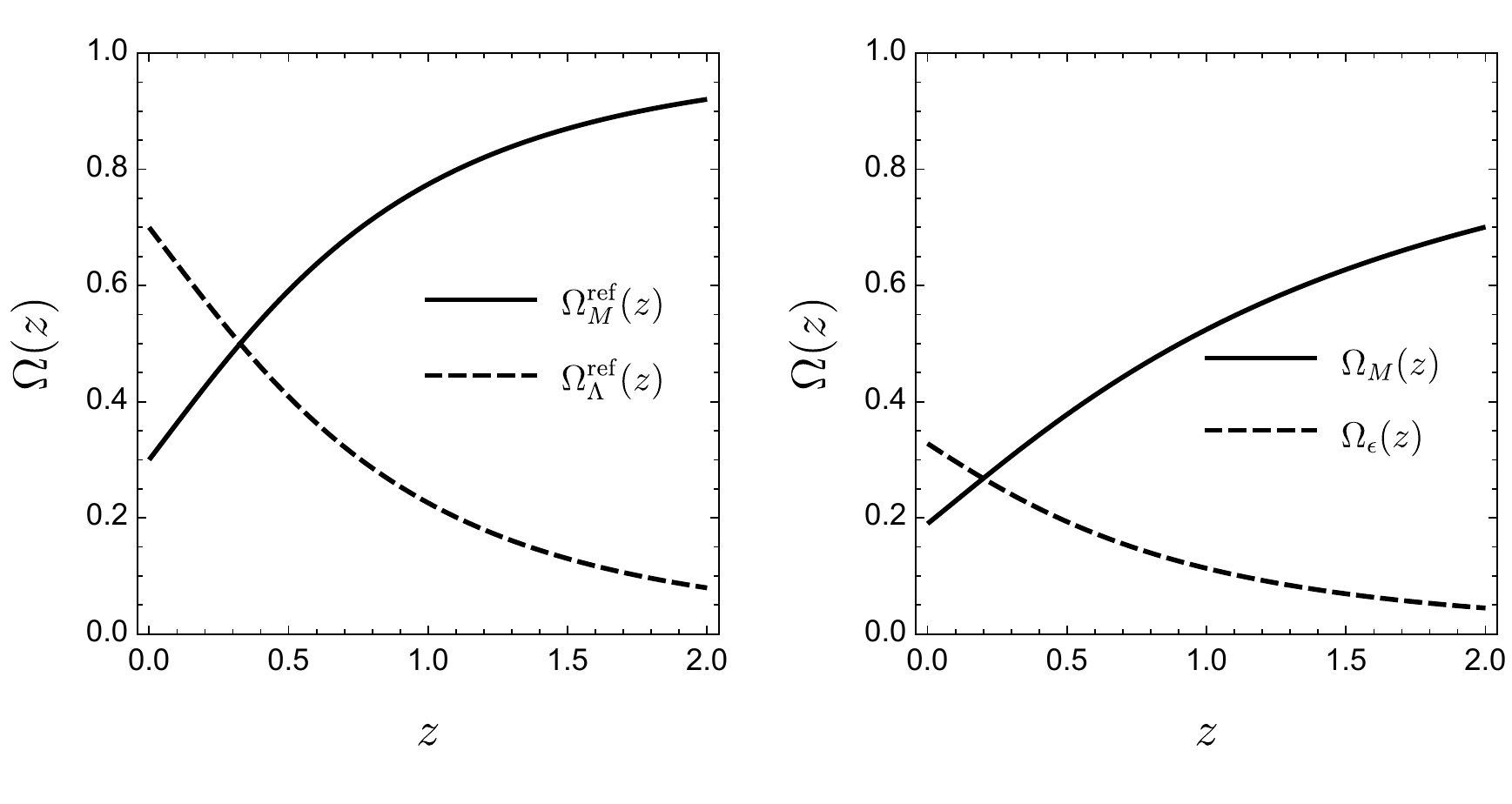}
\caption{This figure shows the evolution of cosmological density parameters as function of
redshift $z$. On the left side, the solid and dashed curves correspond to $\Omega_M^{ref}(z)$ and $\Omega_\Lambda^{ref}(z)$ respectively, for the $\Lambda$CDM model with initial values at $z=0$ given by $\Omega_{M0}^{ref}=0.3$ and $\Omega_{\Lambda0}^{ref}=0.7$. On the right side, the solid and dashed lines correspond to $\Omega_M(z)$ and $\Omega_\epsilon(z)$ respectively, of our model. The considered initial values are $\Omega_{M0}=0.1942$ and $\Omega_{\epsilon 0}=0.3246$. Both plots correspond to a flat Universe}\label{figomegas}
\end{center}
\end{figure}

\begin{figure}
\begin{center}
\includegraphics[scale=0.5]{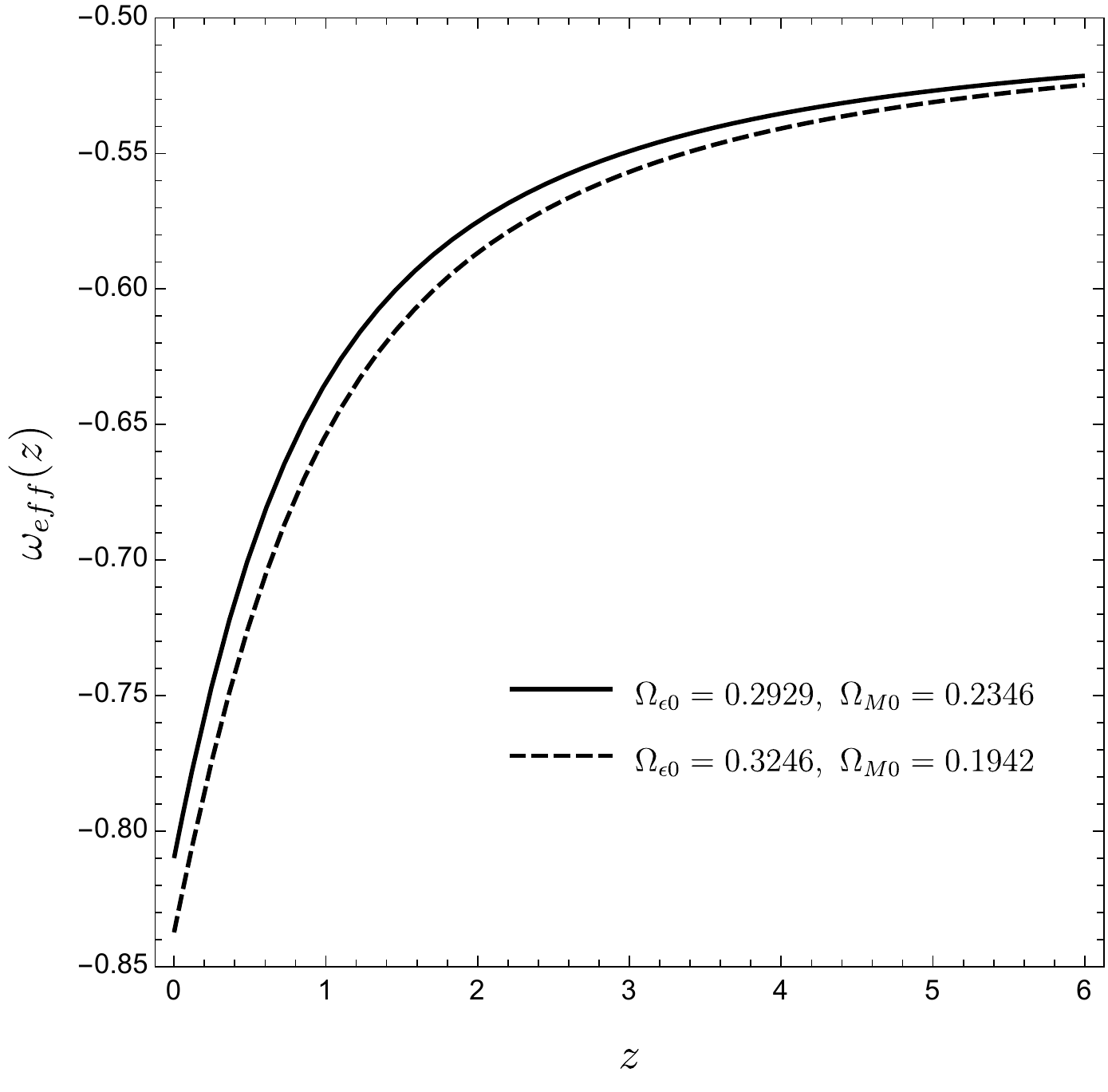}
\caption{Plot of $\omega_{eff}(z)$ obtained  in our model. As seen from the plot, $\omega_{eff}(z)\to -\frac{1}{2}$ for large redshift. The solid line corresponds to $\Omega_{M0}=0.2346$ where $\omega_{eff}=-0.8099$ at $z=0$ while the dashed curve is for $\Omega_{M0}=0.1942$ and $\omega_{eff}=-0.8373$ at $z=0$.}\label{weff}
\end{center}
\end{figure}

{
The $z-$dependent density parameters $\Omega_{M}$ and $\Omega_{\epsilon}$  can be obtained in a straightforward manner, resulting in}
\begin{equation}\label{Omegam}
    \Omega _{M}(z)=\frac{\Omega _{M 0}(1+z)^{3}}{\Omega_{\kappa 0}(1+z)^{2}+ \frac{1}{2}\left[\sqrt{\Omega_{\epsilon 0}}+\sqrt{\Omega_{\epsilon 0}+2\Omega_{M 0}(1+z)^{3}}\right]^{2}},
\end{equation}
and
\begin{equation}\label{Omegaeps}
    \Omega _{\epsilon}(z)=\frac{\Omega _{\epsilon 0}}{\Omega_{\kappa 0}(1+z)^{2}+ \frac{1}{2}\left[\sqrt{\Omega_{\epsilon 0}}+\sqrt{\Omega_{\epsilon 0}+2\Omega_{M 0}(1+z)^{3}}\right]^{2}}.
\end{equation}
In Fig.(\ref{figomegas}) we show the comparison between the behaviour of the density parameters of our model Eq.(\ref{Omegam}) and Eq.(\ref{Omegaeps}) and the behaviour of the density parameters of $\Lambda$CDM, all for a flat universe ($\kappa=0$). 
As seen from all the previous results, our model exhibits the behaviour of a cosmological model with a certain dark energy component. In fact, we can match the model with a standard dark energy model with a specific $z$-dependent parameter of state, as will be detailed bellow.
For a general case, we can consider a $z-$dependent dark energy density parameter as
\begin{equation}\label{omde}
    \Omega_{DE}(z)= \Omega_{X 0}\exp{\left[\int_{0}^{z}3(1+\omega_{X}(z'))d \ln (1+z')\right] },
\end{equation}
and therefore, the Friedmann equation can be written as
\begin{equation}\label{general}
     H^{2}(z)=H_{0}^{2}\left\{\Omega_{\kappa 0}(1+z)^{2}+\Omega_{M 0}(1+z)^{3}+\Omega_{X 0}\exp{\left[\int_{0}^{z}3(1+\omega_{X}(z'))d \ln (1+z')\right] } \right\},
\end{equation}
note that in the case $\omega_{X}=\textrm{constant}$ equation Eq.(\ref{omde}) becomes
\begin{equation}
    \Omega_{DE}(z)=\Omega_{X 0}(1+z)^{3(1+\omega_{X})},
\end{equation}
and we recover Eq.(\ref{usual}). Moreover, as mentioned before, $\omega_{X}=-1$ corresponds to a cosmological constant. For our model, we can find an effective parameter of state $\omega_{eff}(z)$ by requiring Eq.(\ref{general}) to become Eq.(\ref{Hentropic}), this yields 
\begin{equation}
\omega_{eff}(z)=-\frac{1}{2}\left(1+\sqrt{\frac{\Omega_{\epsilon 0}}{{\Omega_{\epsilon 0}+2\Omega_{M0}\left(1+z\right)^3}}}~\right).
\end{equation}
We see that $\omega_{ eff}(z)\to -1$ as $\Omega_{M0}\to 0$, which means that, in order for our model to reproduce the cosmological constant behaviour, the matter density must tend to zero. This agrees with the limit we considered in the previous section to obtain the effective cosmological constant. Also, for larges redshift $\omega_{ eff}(z)\to -1/2$, this limit is independent of the initial values of the density parameters. We present the plot of  $\omega_{ eff}(z)$ in Fig.~(\ref{weff}).

\section{Discussion and Final Remarks}\label{conclusiones}

In the previous section, we showed the viability of the model, but simply adding the volumetric term
on the entropy can be a bit unsettling. As already stated at the beginning, one of the goals of this paper is to establish a physical system that has a volumetric contribution to the entropy-area relationship. It seems plausible that in order to have a volumetric contribution, one might need an extra dimension. In fact, it was shown in \cite{Sheykhi:2007zp} that the entropy for the DGP model has two contributions, one corresponding to gravity in the brane and a second term related to gravity in the bulk. Therefore, we will establish  
a connection of the entropic model (with the volumetric term) with a brane world model.\\
Let us start with a five dimensional Universe, with a three-brane \cite{deffayet,dvali}, this can be described by the action
\begin{equation}
S=\frac{M^3_{(5)}}{2}\int d^5 X\sqrt{\tilde{g}}\tilde{R}+\frac{M^2_{pl}}{2}\int d^4 x\sqrt{g}R,
\end{equation}
where $M_{(5)}$ and $M_{pl}$ are the five and four dimensional Planck masses, respectively. The five dimensional metric is denoted by $\tilde{g}_{A B}$, the brane is located at $y=0$ and has an induced metric given by $g_{\mu\nu}=\tilde{g}_{\mu\nu}(x,y=0)$. The four dimensional Ricci scalar is $R$ and the matter fields are confined to the brane.\\
The line element for the cosmology on the brane \cite{deffayet} is
\begin{equation}
ds^2=-N^2(t,y)dt^2+A^2(t,y)\gamma_{ij}dx^idx^j+B^2(t,y)d y^2,
\end{equation}
the components of the metric are $N(t,y)=1+\varepsilon\vert y\vert\ddot{a}\sqrt{\dot{a}^2+\kappa}$, $A(t,y)=a+\epsilon\vert y\vert\sqrt{\dot{a}^2+\kappa}$ and $B(t,y)=1$. Moreover,  $\gamma_{ij}$ is the three dimensional metric, $\varepsilon=\pm 1$, and $a(t)$ is the scale factor. To get the geometry on the 4D brane, we set $y=0$ and one gets the usual FRW metric from one derives the
the Friedmann equation 
\begin{equation}
H^2+\frac{\kappa}{a^2}=\left( \sqrt{\frac{\rho M^2_{pl}}{3}+\frac{1}{4}r^2_c}+\varepsilon\frac{1}{2r_c}\right)^2,
\end{equation}
where $r_c=\frac{M^{2}_{pl}}{2M^3_{(5)}}$, is the distance scale of the model. {Using the standard continuity equation and } introducing the different matter components with the corresponding equation of state, one defines
\begin{equation}
\Omega_{\alpha 0}=\frac{\rho^0_\alpha}{3M^2_{pl}H^2_0a_0^{3(1+w_\alpha)}},~~~\Omega_{\kappa 0}=\frac{-\kappa}{H^2_0a_0^{2}},~~~\Omega_{r_{c} 0}=\frac{1}{4r^2_cH^2_0}.
\end{equation}
In particular, for non-relativistic matter 
\begin{equation}
H^2(z)=H_0^2\left\{ \Omega_{\kappa 0}(1+z)^2+\left[ \sqrt{\Omega_{r_c 0}}+\sqrt{\Omega_{r_c 0}  +\Omega_{M 0}(1+z)^3}\right]^2 \right \}.
\end{equation}
This is equivalent to Eq.(\ref{Hentropic}). If we identify $\Omega_{r_c 0}=\Omega_{\epsilon 0}/2$, and take $z=0$ we recover the constraint Eq.(\ref{omegaentropic}).
Since we can derive all the cosmological implications from the Friedmann equations and the continuity equation, and taking into account that the Friedmann equations  for the brane world model are the same as the ones
derived from the modify entropy-area relationship, 
we can 
regard both models as equivalent and 
this 
allows us to interpret geometrically the volumetric corrections of Eq.(\ref{entropy}).
Comparing with the DGP model the parameter of the volumetric term on the entropy-area relationship is related to the DGP distance scale  by $\epsilon\sim r^{-1}_c$.

In  this work we have conjectured that the cosmic acceleration has an entropic origin when considering gravity an entropic phenomenon. The main ingredient is a modified entropy-area relationship that includes volumetric contributions. We studied the phenomenological implications of this model and established the viability of this hypothesis. 
Although the modified Friedmann equations gives an effective cosmological constant in the limit $t\to \infty$ and $\rho\to 0$, this models differs from $\Lambda{CDM}$. The best fit gives the two set of values $\left(\Omega_{M 0},\Omega_{\epsilon 0}\right)=\left(0.1942,0.3246\right)$ and $\left(\Omega_{M 0},\Omega_{\epsilon 0}\right)=\left(0.2346,0.2929\right)$. This gives the ranges for the density parameters, $0.2929\le\Omega_{\epsilon 0}\le 0.3246$ and $0.1942\le \Omega_{M0}\le 0.2346$. Furthermore, this {\it ``dark energy''} can be modeled from a barotropic fluid with an $\omega_{eff}(z)$ and for large redshift $\omega_{eff}\to -1/2$. Interestingly, only in the limit $\Omega_{M0}\to 0$ we see that $\omega_{eff}(z)=-1$.\\
Also, we found a relationship between the entropic and DGP  models, that allows us to give a geometrical interpretation to the parameter $\epsilon$. Therefore, this establishes  a physical motivation to the volumetric term in the entropy-area relationship. 
In summary, we can conclude that volumetric contributions on the entropy-area relationship originate the late time acceleration of the Universe and can therefore encode the dark energy sector.

\section*{Acknowledgements}

{\bf J.C.L-D.} is supported by the CONACyT program  ``Apoyos complementarios para estancias sab\'aticas vinculadas a la consolidaci\'on de grupos de investigaci\'on 2022-1'' and by UAZ-2021-38339 Grant.  {\bf I. D. S} supported by the CONACyT program  ``Estancias posdoctorales por M\'exico''. {\bf M. S.} is supported by the grant CIIC 2024.


\end{document}